# High photon flux table-top coherent extreme ultraviolet source

Steffen Hädrich[1,2,*], Arno Klenke[1,2], Jan Rothhardt[1,2,] Manuel Krebs[1], Armin Hoffmann[1], Oleg Pronin[3], Vladimir Pervak[3], Jens Limpert[1,2,4], Andreas Tünnermann[1,2,4]

[1]Friedrich-Schiller-Universität Jena, Abbe Center of Photonics, Institute of Applied Physics, Albert-Einstein-Straße 15, 07745 Jena, Germany

[2]Helmholtz-Institute Jena, Fröbelstieg 3, 07743 Jena, Germany

[3]Ludwig-Maximilian-Universität München, Am Coulombwall 1, 85748 Garching, Germany

[4]Fraunhofer Institute for Applied Optics and Precision Engineering, Albert-Einstein-Straße 7, 07745 Jena, Germany

*email: steffen.haedrich@uni-jena.de

**High harmonic generation (HHG) enables extreme ultraviolet radiation with table-top setups[1]. Its exceptional properties, such as coherence and (sub)-femtosecond pulse durations, have led to a diversity of applications[1]. Some of these require a high photon flux and megahertz repetition rates, e.g. to avoid space charge effects in photoelectron spectroscopy[2–4]. To date this has only been achieved with enhancement cavities[5]. Here, we establish a novel route towards powerful HHG sources. By achieving phase-matched HHG of a megahertz fibre laser we generate a broad plateau (25 eV - 40 eV) of strong harmonics, each containing more than $10^{12}$ photons/s, which constitutes an increase by more than one order of magnitude in that wavelength range[6–8]. The strongest harmonic (H25, 30 eV) has an average power of 143 $\mu$W ($3 \cdot 10^{13}$ photons/s). This concept will greatly advance and facilitate applications in photoelectron or coincidence spectroscopy[9], coherent diffractive imaging[10] or (multidimensional) surface science[2].**

In the late 1980s first experiments on high harmonic generation (HHG) driven by intense pulsed lasers were performed[11,12]. Rapidly, the process attracted significant attention due to its non-perturbative behaviour[12], its coherence[13], the potential of sub-femtosecond pulse trains[14] or isolated attosecond pulses[15] and its table-top setup[12]. Nowadays, significant effort for further advancing this still growing field results in an increasing demand in novel laser sources. The established laser technology for HHG is based on Titanium:Sapphire (TiSa) chirped pulse amplifiers that deliver multi-millijoule ultrashort (~25 fs) pulses with only several kilohertz of repetition rate at 800 nm. An increase in repetition rate to the megahertz level with an increased photon flux, i.e. the number of photons per second per harmonic, would e.g. help to mitigate space charge effects in photoelectron spectroscopy[2,3], reduce integration times in coherent diffractive imaging[10] or increase the signal-to-noise ratio in general.

Inherently, increasing the repetition rate in HHG is associated with average power scaling of ultrashort-pulse lasers, which is challenging to realize. In that regard, the use of passive enhancement cavities (EC), which coherently overlap pulses inside a high finesse resonator, has been considered as most promising approach so far[5,16]. HHG is directly achieved inside the resonator, which requires an out-coupling mechanism by default. The latter one and ionisation-induced phase-shifts have been identified as most severe limitations[17]. Yet, the highest average power of approximately 200 $\mu$W in the 11$^{th}$ harmonic (12.7 eV) of an enhanced 1070 nm fibre lasers has been obtained with this approach, however, with rapidly decreasing signal towards 30 eV[5]. Recently, the use of ECs was extended to shorter wavelengths by demonstrating harmonics up to 100 eV and 5.3 $\mu$W in the 27$^{th}$ harmonic (32.5 eV)[6].

At the same time, the complementary approach of directly using high repetition rate lasers for HHG has been pursued[18]. Although up to 20 MHz were demonstrated, the attainable photon flux was orders of magnitude lower than that of ECs or traditional TiSa lasers, because phase matching was not achieved[19].

In this work, we demonstrate phase matched HHG of a nonlinear compressed fibre chirped pulse amplifier resulting in more than 10$^{-6}$ conversion efficiency into a single harmonic at 30 eV. This demonstrates a new class of coherent extreme ultraviolet sources that increases the average power in the 25 eV-40 eV range by more than one order of magnitude over previous systems[6–8].

High harmonic generation can be understood with a simple three-step model that describes the response of a single atom to a strong laser field[20]. First, an electron is tunnel-ionized, then it is accelerated in the laser field and, finally, recombines with the parent ion emitting a photon that carries the acquired kinetic plus the binding energy. The single atom response to a certain intensity, i.e. the dipole amplitude $A_q$, can be obtained quantum mechanically[21] or modelled by empirical scaling laws[22]. The overall yield obtained in HHG, however, critically depends on macroscopic effects (phase matching), i.e. the coherent build-up along the generation medium. The dipole amplitude $A_q$ significantly increases with intensity, but at the same time the increased ionisation fraction strongly influences the phase matching[1,23]. It has to be noted that the use of ultrashort laser pulses (<10 cycles) reduces the impact of ionization and allows using higher intensities[1]; therefore, increasing $A_q$. Ultimately, the signal build-up is limited by linear absorption of the harmonics in the generation medium itself[24].

For the experiments presented, we use a fiber chirped pulse amplifier with subsequent nonlinear compression to achieve the required ultrashort pulses (see Methods). The system delivers laser pulses with 130 µJ-150 µJ of energy, a duration of approximately 30 fs (Fig.1a) and variable repetition rates[25]. These pulses are focused to a focal spot diameter of 90 µm ($1/e^2$ intensity, Fig. 1b) inside a vacuum chamber that contains gas jets (Fig.1c) of different diameter. The generated harmonics and the remaining infrared light pass two $SiO_2$ surfaces (see Methods) reducing the average power of the driving laser to protect the following two aluminium filters (1 µm thickness) against damage. Subsequently, the harmonics are analysed with a flat-field grating based spectrometer (see Methods). The optimization of the experimental parameters has been carried out for xenon gas, since it has the lowest ionization potential and highest $A_q$ of all noble gases. The intensity in the focal spot has been gradually reduced from $1.5 \cdot 10^{14}$ W/cm$^2$ until blue shifting of the harmonics with increasing pressure disappeared at $\sim 9 \cdot 10^{13}$ W/cm$^2$. Figure 2a shows the signal of the 23$^{rd}$, 25$^{th}$ and 27$^{th}$ harmonic, obtained at this intensity level, with respect to various nozzle openings. Obviously, the signal is increasing up to the maximum opening diameter of 1 mm, where it starts to saturate. This behaviour is very similar for all observed harmonics (H19-H31) leading to broad plateau of strong harmonics. The absolute signal of the harmonics is obtained either with the known detection efficiency or with a photodiode (see Methods). The latter method was used to measure the average power of three harmonics (Fig.2b) at 50 kHz. This measurement yields 10.7 µW (H19), 13.3 µW (H21) and 15.5 µW (H23), which is in good agreement with our estimation, based on known detection efficiency, of 8 µW (H19), 10 µW (H21) and 13 µW (H23), respectively. Due to the consistency of both techniques the photon flux for the remaining harmonics and higher repetition rates has been obtained by using the detection efficiencies. Additionally, we have studied the phase matching behaviour by varying the backing pressure of the gas jet and, simultaneously, recording the signal of the respective harmonics. The obtained signal growth of the harmonics is also in good agreement with a numerical model[22,24] (Fig.2 c, see Supplement). According to a calculation with the ADK model the ionization fraction at the peak of the pulse is 24 % and higher than the critical ionization level[23]. Therefore, the experiment is performed in a transient phase-matching regime, where the coherence length $L_{coh}=\pi/\Delta k$ is larger than the medium length over a short time interval at the rising edge of the pulse where the harmonics are generated and phase matching is achieved. The absorption length at optimal pressure (~60 mbar in the interaction region, Fig. 2c) is between 430 $\mu$m (H21) and 2.4 mm (H31), which means that the lower order harmonics are close to the absorption limit, while a longer interaction length would be required for higher orders[24] (see Supplement). More importantly, the repetition rate of the laser system can be increased up to 0.6 MHz (80 W of average power) with very similar pulse parameters, which is corroborated by the almost linear increase of the harmonic signal (inset of Fig.2b). The spatial profiles and the spatially integrated and calibrated spectra for the experiments at 0.6 MHz are shown in Figure 3a. The gas jet is positioned slightly behind the focus (~150 $\mu$m), which results in phase matching for the short trajectories as indicated by the excellent spatial profiles of the low diverging harmonics[13]. A similar optimisation has been carried out for krypton gas targets (Fig. 2d) resulting in the use of a 600 $\mu$m nozzle placed ~180 $\mu$m behind the focus. The intensity of $10^{14}$ W/cm$^2$ (50 kHz) and $8.8 \cdot 10^{13}$ W/cm$^2$ (600 kHz) is similar, but the ionisation is significantly reduced to <3 % at the pulse peak. Consequently, phase matching can be achieved over an increased time window around the pulse peak, which results in spectrally narrower harmonics (Fig.3b). In this case the coherence length is ~1 mm and the absorption length for the optimal pressure of 110 mbar is between 162 $\mu$m (H23) and 730 $\mu$m (H33), which means that again the lowest harmonic orders are generated absorption-limited (see Supplement). The obtained average power of the harmonics generated in xenon and krypton are shown in the lower panels of Figure 3. Strong harmonics with more than 30 $\mu$W per harmonic, i.e. more than $10^{12}$ photons/s, are obtained over a broad plateau extending from the 21$^{st}$ to the 33$^{rd}$ harmonic (25 eV – 40 eV). The highest average power is obtained for H25 (30 eV) with 143 $\mu$W, which corresponds to $3 \cdot 10^{13}$ photons/s and a conversion efficiency of $1.8 \cdot 10^{-6}$. The achieved photon flux is already within one order of magnitude of large-scale facilities, such as free electron lasers[26].

In conclusion, we have demonstrated the most powerful source of coherent extreme ultraviolet radiation (25 eV – 40 eV) enabled via high harmonic generation to date[6–8]. The combination of megahertz level repetition rate and highest photon flux will greatly advance applications in various fields, in particular, in (multi)-dimensional surface science, coincidence detection experiments, photoelectron emission spectroscopy (PES) and microscopy (PEEM), (time-resolved) coherent diffractive imaging (CDI) or others[2,4,9,10]. Due to the excellent scaling properties of coherent combination the availability of kilowatt average power femtosecond lasers[27] will further increase the HHG signal by another order of magnitude. Moreover, a second nonlinear compression stage can be implemented[28] for achieving, potentially, sub-10 fs pulses with high average power and repetition rate. Experiments performed at 918 nm[29] have shown that such pulses should enable to address the soft x-ray or even water window region that has particular importance for biological imaging. In combination with carrier-envelope phase stabilisation this could also enable high photon flux megahertz isolated attosecond pulses. Therefore, the presented results are a major milestone towards new applications of coherent extreme ultraviolet radiation in science and technology.

**Methods**

*Fibre laser with nonlinear compression*

The front-end is a fibre chirped pulse amplification (FCPA) system that incorporates coherent combination (CC) of pulses from up to four main amplifier channels[30]. For the experiments presented here, the system is operated with 270 µJ pulse energy, a compressed pulse duration of approximately 340 fs and a repetition rate between 50 kHz and 0.6 MHz, which corresponds to an average power between 14 W and 163 W. The laser pulses are sent to the nonlinear compression setup that is described in[25]. They are coupled into the 1.1 m long hollow-core fiber with an inner diameter of 250 µm. After evacuation the tube is filled with 4 bar of krypton gas to enable spectral broadening. Subsequently, the pulses are compressed in time by a chirped mirror compressor with a group delay dispersion of -1600 fs$^2$. After propagation through this setup the pulses are 29 fs short (Fig.2b)) and have an energy of 130 µJ-150 µJ. At the highest repetition rate of 600 kHz the average power used for high harmonic generation is >80 W.

*High harmonic generation and characterization of extreme ultraviolet radiation*

The nonlinear compressed pulses are focused with an f=300 mm lens (~f/50 focusing) onto a gaseous target inside a vacuum chamber. Simple cylindrical nozzles with different opening diameters (see text) provide the gas targets. The generated harmonics and the fundamental infrared laser co-propagate and impinge on a chicane of two SiO$_2$ substrates, which are used with 75° angle of incidence and p-polarisation. The first substrate is pure fused silica, while the second one contains an anti-reflection coating for the infrared with the top layer being SiO$_2$[31]. Consequently, the surfaces reflect the harmonics with sufficient efficiency (e.g. ~17 % at 30 eV for two reflections), while the infrared is suppressed to less than 1%. This allows the following two aluminium filters to withstand even in high average power operation. After the filters the infrared light is completely suppressed and the harmonics are sent into a flat-field grating based spectrometer (Ultrafast Innovations GmbH) that is equipped with a charge coupled device (CCD) camera (Andor, Newton).

We have used two independent methods to obtain the average power of individual harmonics. As described in the main section the first method relies on filtering 3 harmonics (Fig. 2b) by using a combination of a 200 nm aluminium and a 200 nm zirconium filter and then measuring the average power with a photodiode (AXUV100G, Opto Diode). The measured current of the photodiode is converted into an average power with the known responsivity of the diode of (0.26 ± 0.01) A/W at 23 - 27 eV. This power is distributed among the harmonics according to the percentage values given in Fig.2 b. The harmonic signal before the filters is obtained by using the measured transmission values of each filter and harmonic, respectively. The reflection coefficient of the $SiO_2$ surfaces is calculated using the general reflection coefficients[32] and the tabulated values for the complex refractive index of $SiO_2$[33] in the extreme ultraviolet range, which have shown good agreement with calibration measurements[31].

The second method is based on calculating the harmonic signal by accounting for the known efficiencies of the detection apparatus. The number of photons per second $N_{ph,s}$ emitted directly after the gas jet is obtained by using the following equation

$$N_{ph,s} = \frac{S_{CCD} \cdot \sigma}{\eta_{QE} \cdot (E_{ph}/3.65eV) \cdot \eta_g \cdot t_{f1} \cdot t_{f2} \cdot R_p^2 \cdot t_{exp}},$$

where $S_{CCD}$ is the measured signal (counts) on the detector, $\sigma$ is the CCD sensitivity in electrons per A/D count, $\eta_{QE}$ is the quantum efficiency of the CCD, $E_{ph}$ is the photon energy of the harmonics in eV ($E_{ph}$/3.65 eV is the number of electrons freed per photon at a bandgap energy of 3.65 eV), $\eta_g$ is the grating diffraction efficiency, $t_{f1/2}$ are the filter transmissions, $R_p$ is the reflection coefficient of a $SiO_2$ surface for p-polarized light and $t_{exp}$ is the exposure time. The CCD characteristics ($\sigma$, $\eta_{QE}$) are used as characterized by the manufacturer, the diffraction efficiency of the grating $\eta_g$ has been used as described in[29], $R_p$ is obtained as described above and the transmission of the aluminium filters is measured. The so obtained values have been compared to the photodiode measurement performed for the harmonic orders H19, H21 and H23 and show good agreement (see text).

*One-dimensional model for high harmonic generation*

We have used a one-dimensional model that calculated the harmonic signal on axis[22,24]. For that purpose we have calculated the time-dependent wave-vector mismatch $\Delta k(t) = q \cdot k_0 - k_q$ according to our experimental conditions. The model proves very useful for understanding the phase matching conditions present in our experiments. More details on the model and experimental conditions are described in the supplement.

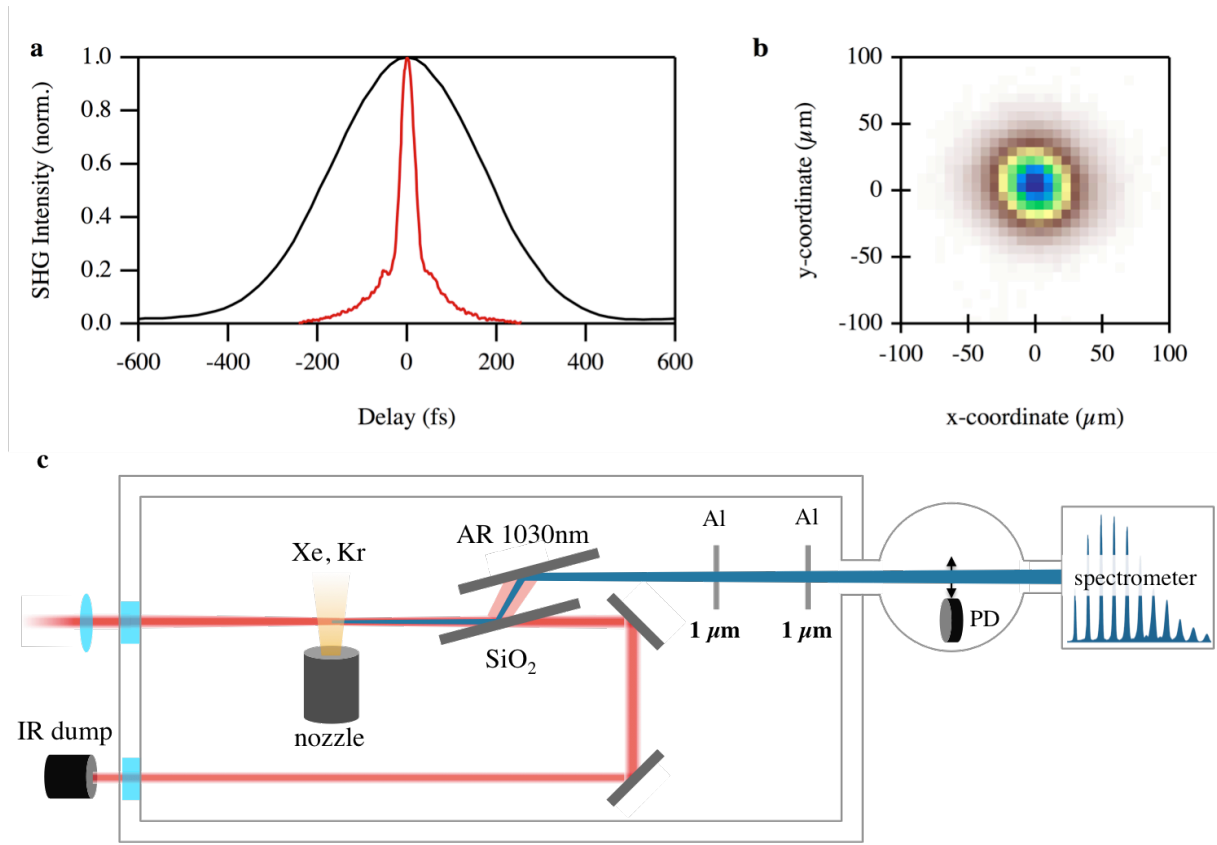

**Figure 1. Experimental setup of the high harmonic generation experiments.**

**a**, The pulses of a fibre chirped pulse amplifier are post-compressed (see Methods). The figure shows the autocorrelation traces of the 340 fs (black) fibre laser and the 30 fs (red) compressed pulses. **b**, The intensity profile of the focal spot has a diameter of 90 $\mu$m (1/e$^2$ intensity). **c,** The experimental setup after the nonlinear compression stage comprises a vacuum chamber used for the experiments on high harmonic generation. The pulses are focused (**b**) into a gaseous target (krypton, xenon) provided by simple cylindrical opening nozzles of various sizes (see text). The generated harmonics co-propagate with the infrared beam and impinge on a chicane of two SiO$_2$ substrates under 75° angle of incidence. Two aluminium filters, each with a thickness of 1 $\mu$m, isolate the harmonics, which are then sent into a flat-field grating spectrometer or onto a photodiode (see Methods).

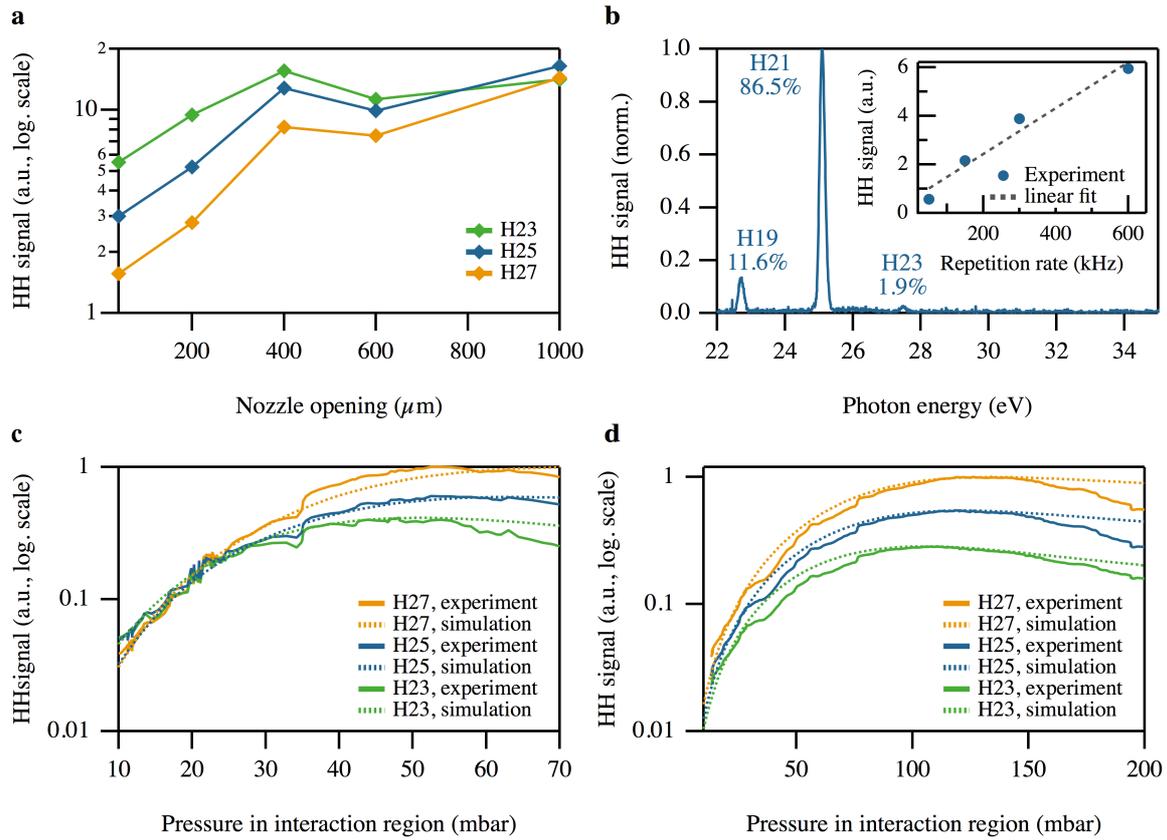

**Figure 2. Optimisation and phase-matching in high harmonic generation.**

**a,** The spatially and spectrally integrated signal of three harmonics (H23, H25 and H27) generated in xenon are shown with respect to the size of the nozzle opening. **b,** The blue curve shows the spectrum of the high harmonics transmitted through a combination of a 200 nm aluminium and 200 nm zirconium filter together with their respective percentage of the overall signal as used for the photodiode measurement (see text and Methods). The inset shows the harmonic signal (sum over H21-H29) as a function of the repetition rate. **c,** The signal of harmonics H23-H27 (solid lines) generated in a 1 mm xenon gas jet has been recorded with respect to the pressure. Additionally, the result of a simulation (dashed lines, see Supplement) is shown for a I=8·10$^{13}$ W/cm$^2$, 30 fs pulse. **d,** The signal of harmonics H23-H27 (solid lines) generated in a 600 µm krypton gas jet has been recorded with respect to the pressure. Additionally, the result of a simulation (dashed lines, see Supplement) is shown for a I=9.7·10$^{13}$ W/cm$^2$, 30 fs pulse.

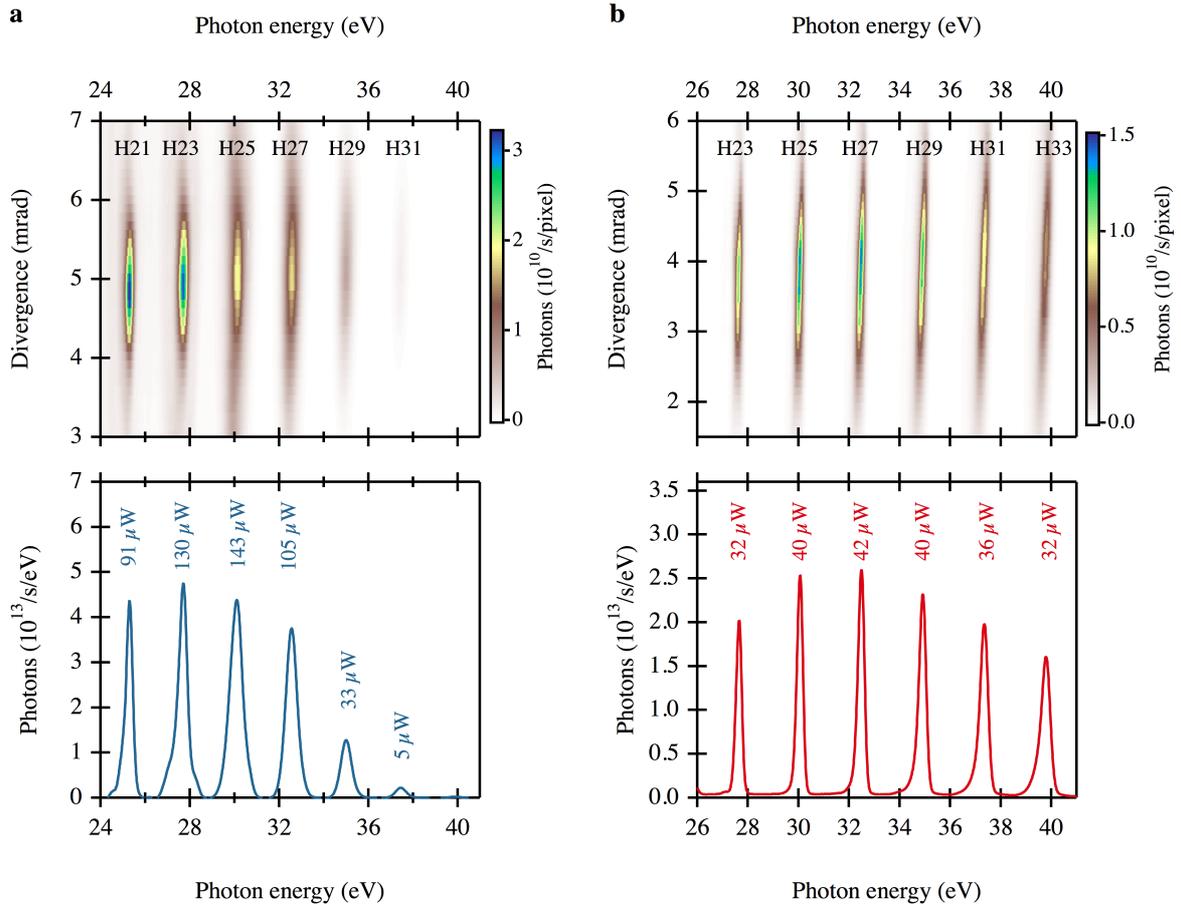

**Figure 3. High harmonic generation at 0.6 MHz repetition rate.**

**a,** The upper panel shows spatial (y-axis) and spectral (x-axis) profiles of the harmonics generated in a 1 mm xenon jet with the laser being operated at 0.6 MHz repetition rate. The lower panel shows the spatial integration of the upper panel together with the corresponding number of photons per second of each harmonic line. The strongest harmonic is H25 (30 eV) with an average power of 143 $\mu$W corresponding to $3\cdot 10^{13}$ photons/s. **b,** The upper panel shows spatial (y-axis) and spectral (x-axis) profiles of the harmonics generated in a 600 $\mu$m krypton jet with the laser being operated at 0.6 MHz repetition rate. The lower panel shows the spatial integration of the upper panel together with the corresponding number of photons per second of each harmonic line. The strongest harmonic is H27 (32.5 eV) with an average power of 42 $\mu$W corresponding to $8\cdot 10^{12}$ photons/s.

**Acknowledgements**

This work has been partly supported by the German Federal Ministry of Education and Research (BMBF) and the European Research Council under the European Union's Seventh Framework Programme (FP7/2007-2013) / ERC Grant agreement n° [240460]. A.K. acknowledges financial support by the Helmholtz-Institute Jena.

**Author Contributions**

J.L., S.H., J.R. and M.K. conceived the experiment. The experiments were planned and performed by S.H., J.R., A.K., A.H. and M.K. The data were analysed by S.H. with support from J.R. and M.K. All authors discussed and contributed to the interpretation of the results. J.L. and A.T. supervised the project and acquired funding. The idea and design of the anti-reflection coated $SiO_2$ substrates originate from O.P. and V.P., which also fabricated the samples used in this experiment. All authors contributed to the manuscript.

**Competing Financial Interests**

The authors declare no competing financial interest.